\documentclass[twocolumn,
amsmath,amssymb]{revtex4}
\usepackage{amssymb}
\usepackage[dvips]{graphicx}
\begin{document}

\title{Quantum phase transitions in dilute bosonic superfluids on a lattice}
\author{A. S. Alexandrov and I. O. Thomas}

\affiliation{Department of Physics, Loughborough University,
Loughborough LE11 3TU, United Kingdom\\}

\begin{abstract}It has  been well known that quantum fluctuations induce a macroscopic phase transition from
a superfluid to a Mott insulator phase driven by the repulsive
potential energy in the ground state of dense bosonic systems on a
lattice. We find a quantum phase  transition  from the homogeneous
to an inhomogeneous Bose-condensate strongly affected or sometimes
driven by the kinetic energy dispersion in dilute bosonic
superfluids, which provides a clear identification of the superfluid
state.
\end{abstract}

\pacs{03.75.Hh, 03.75.Lm}
\maketitle

The experimental realization of the quantum phase transition (QPT)
from a superfluid liquid to a Mott insulator phase in an atomic gas
trapped in an optical lattice \cite{greiner}  triggered remarkable
experimental and theoretical activity. This experiment heralded a
new regime in exploration of the many body physics dominated by an
interplay between atom-atom interactions and the boson kinetic
energy as described by the Bose-Hubbard model \cite{jaksch}. The
competition between kinetic and interaction energy terms in the
underlying Hamiltonian has been considered to be fundamental to
quantum phase transitions not only in neutral  but also in charged
Bose liquids in the context of granular superconductors and
Josephson junction arrays \cite{fisher} and preformed real-space
electron pairs as bipolarons \cite{alemot}.

The Bose superfluid to Mott insulator transition takes place only at
fixed density in a homogeneous dense system when the number of
bosons is commensurate with the number of lattice sites. At first
glance, one would not expect any phase transformation in  a dilute
Bose superfluid  well described by the Bogoljubov theory \cite{bog}.
Surprisingly (as we shall show), this wisdom is not always
applicable to bosons in a periodic potential, whose  ground state is
described by a Gross-Pitaevskii-type (GP) equation
\cite{gross,pitaevskii} including lattice, $V({\bf r})$, and
interaction, $U({\bf r})$, potentials,
\begin{equation}
\left[-{\hbar^2 \nabla^2 \over{2m}}+V({\bf r}) -\mu +  \int d{\bf
r'} U({\bf r} -{\bf r'})|\psi ({\bf r'})|^2\right] \psi({\bf r}) =0.
\label{gp}
\end{equation}
Here the condensate wave-functions $ \psi({\bf r})$ , which is also
the order parameter, is normalized by the number of bosons as $\int
d{\bf r} |\psi({\bf r})|^2 =N_b$, $m$ is the boson mass, and $\mu$
is the chemical potential, which controls the total number of
particles in the system.
\begin{figure}
\begin{center}
(a)
\includegraphics[angle=-0,width=0.42\textwidth]{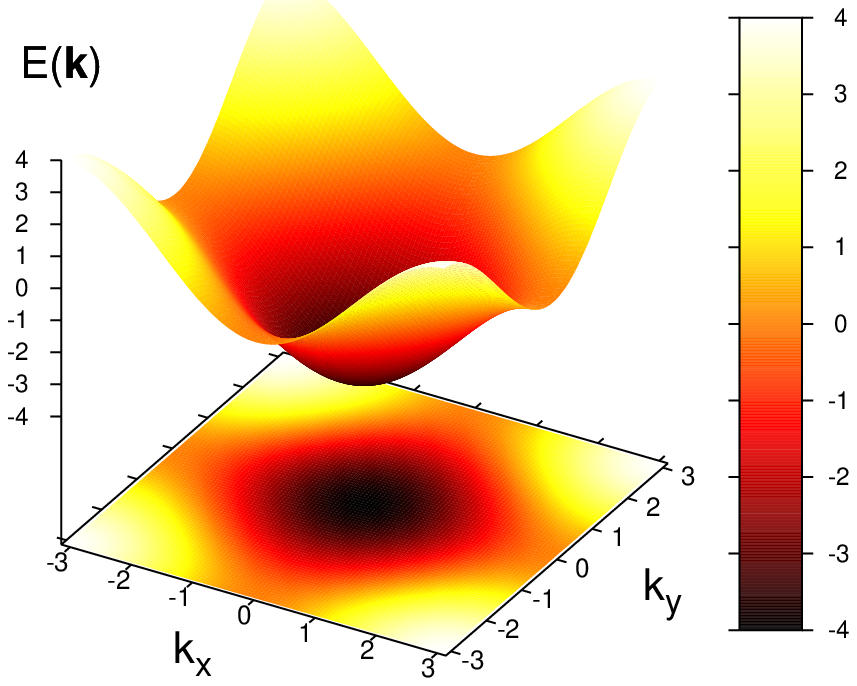}
(b)
\includegraphics[angle=-0,width=0.42\textwidth]{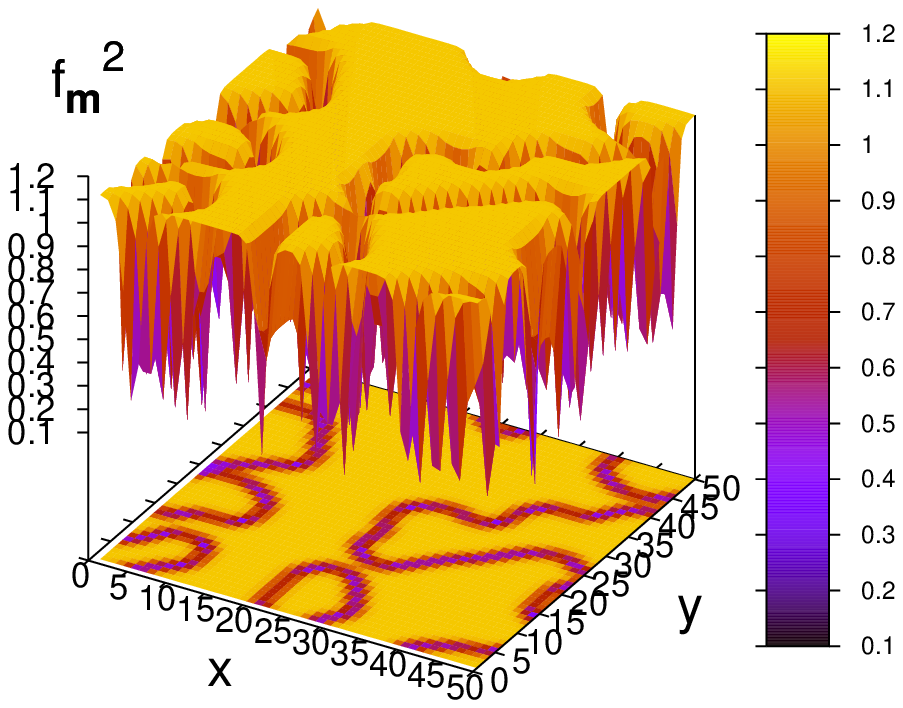}
(c)
\includegraphics[angle=-0,width=0.42\textwidth]{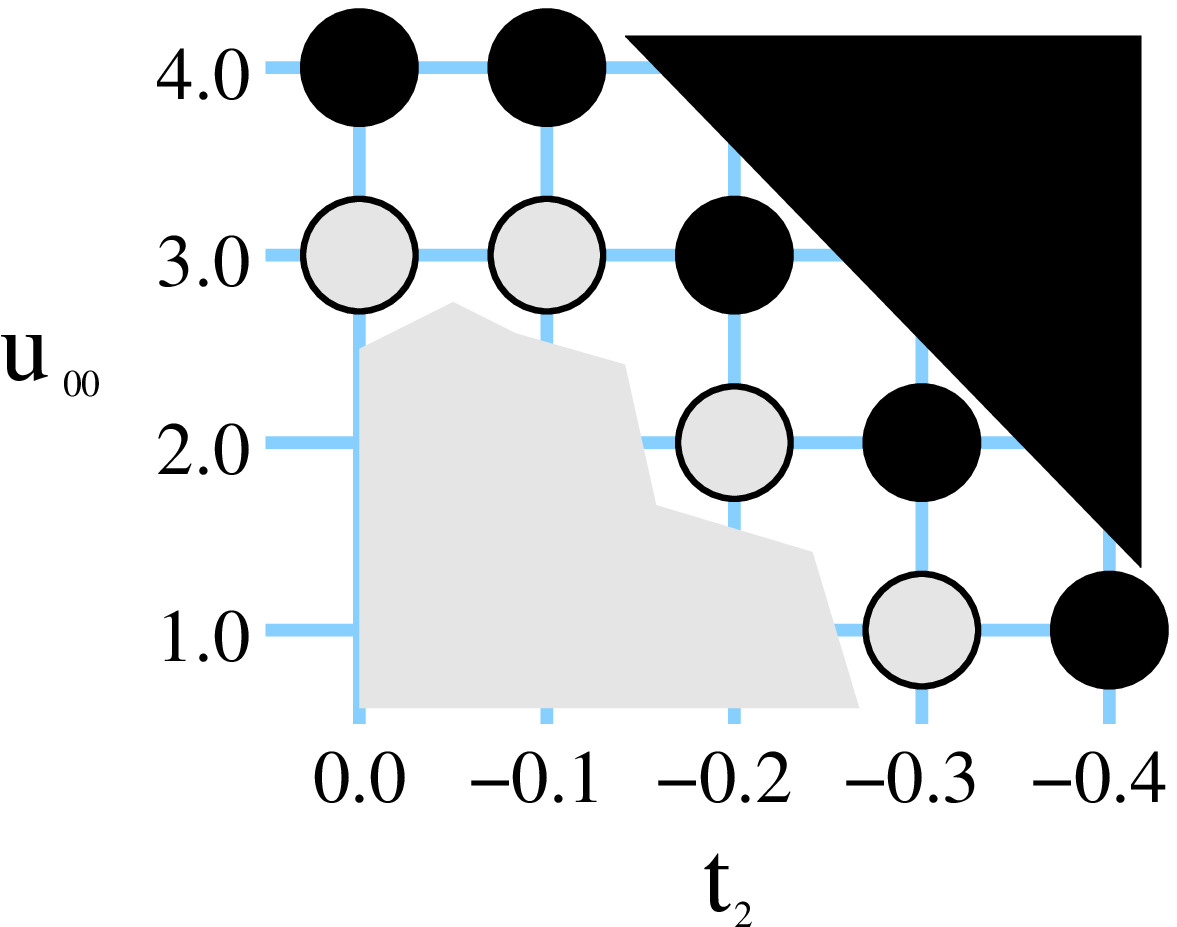}
\vskip -0.5mm \caption{(a) Energy band dispersion and (b)
inhomogeneous density when $u_{00}=4.0$ in a deep square lattice
potential with the hopping integrals $t_1=1,t_2=t_3=0$. c) Location
of onset of inhomogeneity as $t_2$ is varied.  Gray indicates a
homogeneous condensate, black inhomogeneous. Filled circles
correspond to the locations of  simulations run on $50\times 50$
lattices in the border region.}\label{fig:1}
\end{center}
\end{figure}
One can solve equation (\ref{gp}) variationally on a representative
lattice with a large number of sites, $N$,  using a complete set of
orthogonal Wannier (site) functions $w({\bf r})$. Transforming the
order parameter as  $ \psi({\bf r})= \sum_{\bf m} \phi _{\bf m} w
({\bf r-m})$ we reduce (\ref{gp}) to a discrete set of equations for
the site amplitudes $\phi _{\bf m}$,
\begin{equation}
-\sum_{\bf m} [t({\bf m-m'})+\mu \delta_{\bf m,m'}] \phi _{\bf m} +
\sum_{\bf m, n,n'}U_{\bf mn}^{\bf m'n'} \phi^* _{\bf n'}\phi _{\bf
n}\phi _{\bf m} =0. \label{gpsite}
\end{equation}
Here $t({\bf m})=\int d{\bf r} w^*({\bf r})[-\hbar^2 \nabla^2/2m
+V({\bf r})] w({\bf r-m})$
 is the hopping integral, and  $U_{\bf m n}^{\bf m'n'}=\int \int d{\bf r}d{\bf r'}U({\bf
r-r'}) w^*({\bf r-m'})w^*({\bf r'-n'})w({\bf r'-n})w({\bf r-m})$
 is the matrix element of the interaction potential.

We are interested in the dilute liquid regime far away from the Bose
liquid-Mott transition. It is accustomed to keeping only the
density-density interactions, $U_{\bf mn}^{\bf m'n'}\approx U_{\bf
mn}^{\bf m n}\delta_{\bf m,m'}\delta_{\bf n,n'}$. Solving the GP
equation (\ref{gpsite}) is then equivalent to minimization of the
energy functional,
\begin{equation}
E(\phi _{\bf m})=-\sum_{\bf m,n} \left[t({\bf m-n})+\mu \delta_{\bf
m,n}
 + {1\over{2}}U_{\bf m
n}^{\bf m n} \phi^* _{\bf m} \phi _{\bf n}\right]\phi^* _{\bf n}\phi
_{\bf m}. \label{functional}
\end{equation}
Importantly, rescaling the order parameter as $\phi _{\bf m}=n^{1/2}
f_{\bf m}$ and the interaction as $U_{\bf mn}^{\bf m n}= u_{\bf
nm}/n$ yields a universal functional, $\tilde{E}(f_{\bf m})=E(\phi
_{\bf m})/n$ that is independent of $n=N_b/N$, the number of bosons
per lattice valley:
\begin{equation}
\tilde{E}(f_{\bf m})=-\sum_{\bf m,n} \left[t({\bf m-n})+\mu
\delta_{\bf m,n}
 + {1\over{2}}u_{\bf m
n} f^* _{\bf m} f _{\bf n}\right]f^* _{\bf n}f _{\bf m}.
\label{functional2}
\end{equation}
A solution to the tight-binding form of (\ref{gp}) corresponds to
the minimum of this functional when the following constraint is
imposed, $\sum_{\bf m}|f_{\bf m}|^2=N$.  Solutions for different
particle densities are mapped on to each other by a simple rescaling
of the interaction.

\begin{figure}
\begin{center}
(a)
\includegraphics[angle=-0,width=0.42\textwidth]{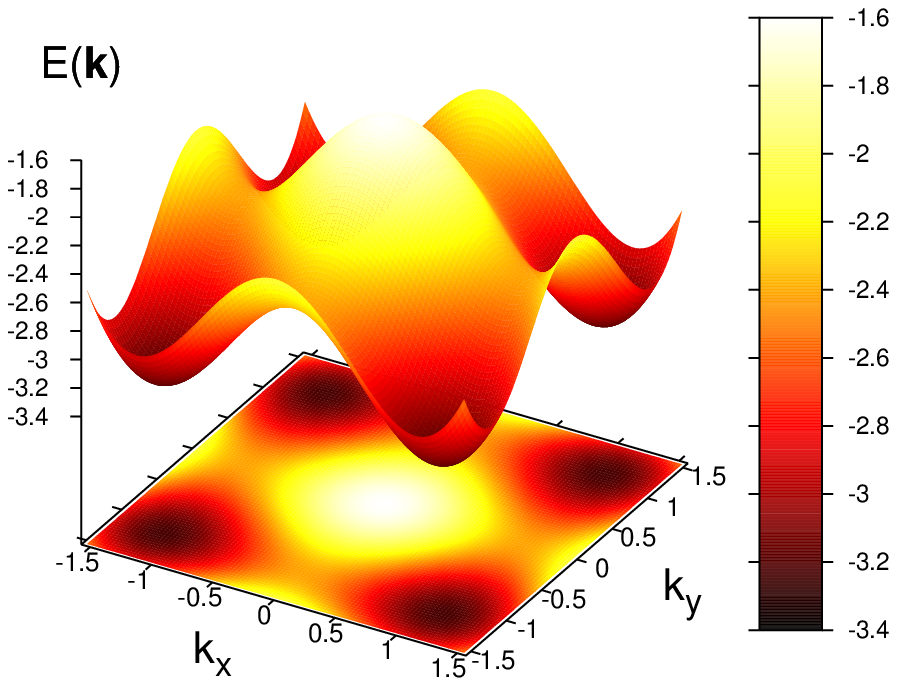}
(b)
\includegraphics[angle=-0,width=0.42\textwidth]{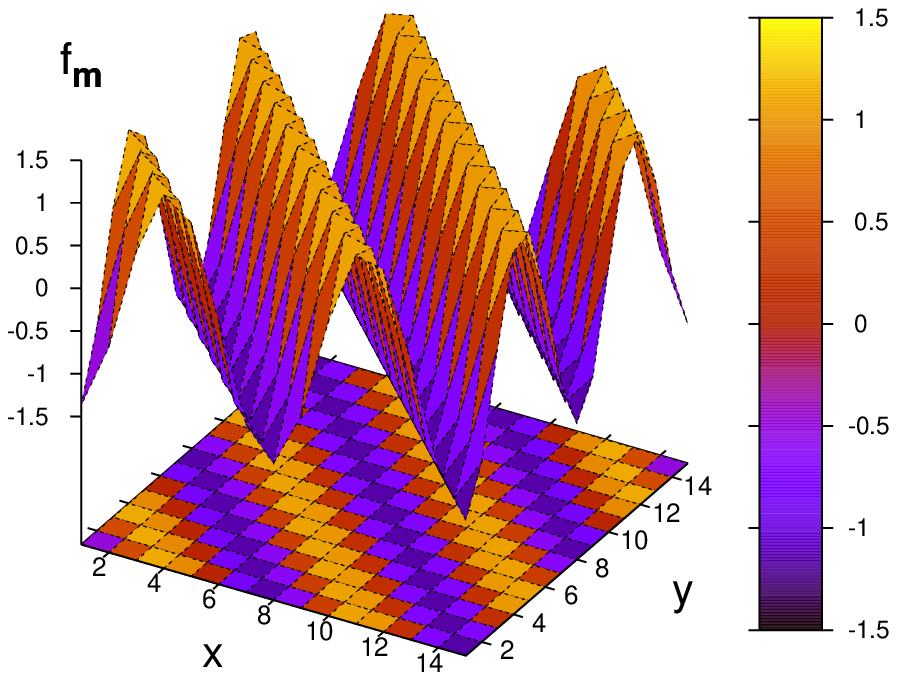}
(c)
\includegraphics[angle=-0,width=0.42\textwidth]{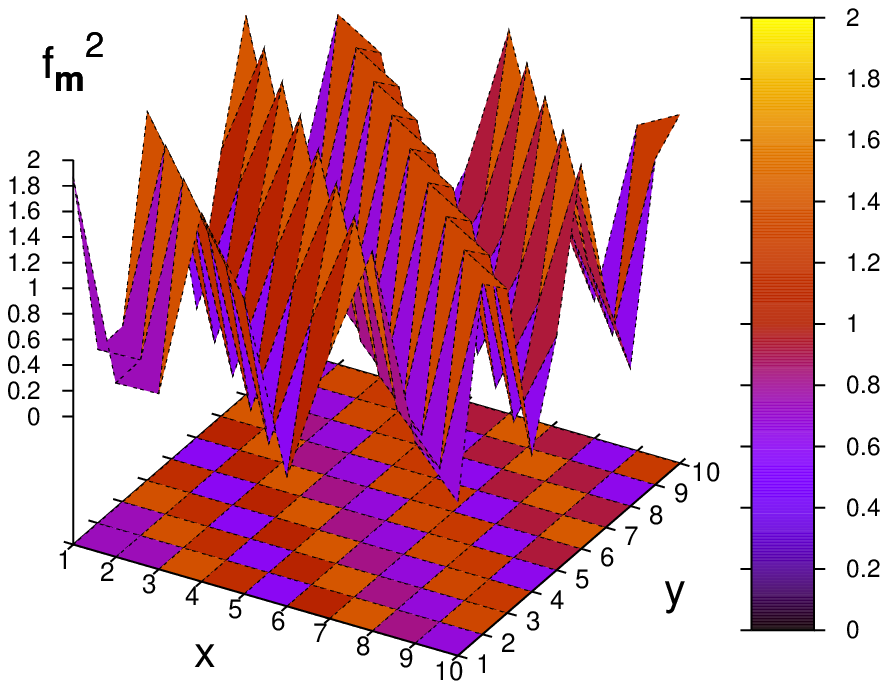}
\vskip -0.5mm \caption{Energy band dispersion in a shallow square
lattice potential with $t_1=1,t_2=0, t_3=-0.4$ (a),  real space
order parameter (b), and  striped particle density (c) on a section of a $50\times50$ lattice. Here the
weak on-site repulsion is  $u_{00}=0.02$ with no inter-site
interactions.} \label{fig:2}
\end{center}
\end{figure}

 The noninteracting part of Eq.(\ref{gp})
can be solved using the Bloch eigenfunctions, and the Wannier states
are constructed by summing the Bloch states with appropriate phase
factors.  Having obtained the Wannier states the hopping integrals
and the interaction matrix elements can be evaluated for any
periodic lattice potential $V({\bf r})$. When the lattice potential
is deep enough, the  nearest-neighbor hopping, $t_1$ dominates over
the next-nearest  (nn), $t_2$ and the next-next (nnn), $t_3$ nearest
neighbor tunneling  \cite{blakie}. Lowering the lattice potential
causes $t_2$ and/or $t_3$ hopping integrals to be of the same order
as $t_1$. Also lowering the boson density reduces the rescaled
repulsion between bosons. That allows us to investigate QPTs in the
superfluid state which might be driven by the kinetic energy
dispersion, $E({\bf k})= \sum_{\bf m} t({\bf m}) \exp(i{\bf k \cdot
m})$ rather than by the repulsion under certain conditions.

We have randomly generated starting values for real site amplitudes,
$f_{\bf m}$. The kinetic portion of the functional is calculated in
the momentum space using the Fourier transformed values of the
amplitudes at each lattice site. Then it is inverse
Fourier-transformed back into the Wannier space, where the effects
of the interaction energy are applied; note that the minimization
procedure tends to fail in the absence of any repulsion. The
functional (\ref{functional2}) is minimized for a given value of
$\mu$ using the NAG optimization routine E04DGF.  The value of $\mu$
is then fine-tuned so that the constraint  $\sum_{\bf m}|f_{\bf
m}|^2=N$ is imposed. There are virtually no significant size effects
as verified  by  our simulations on $N=25 \times 25$ and $N=50\times
50$ lattices.

\begin{figure}
\begin{center}
(a)
\includegraphics[angle=-0,width=0.45\textwidth]{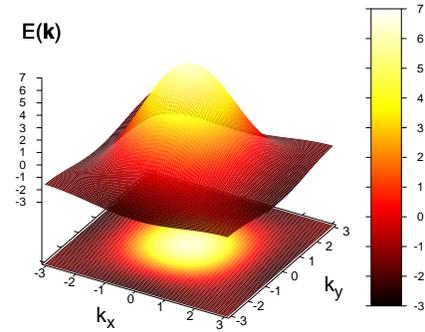}
(b)
\includegraphics[angle=-0,width=0.45\textwidth]{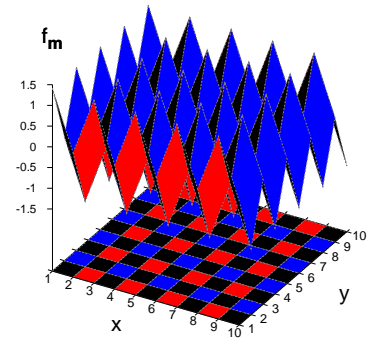}
(c)
\includegraphics[angle=-0,width=0.45\textwidth]{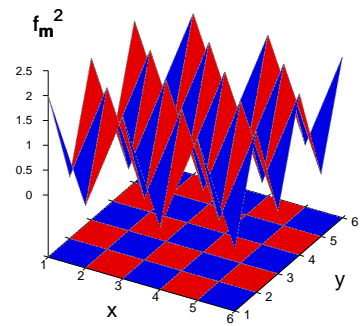}
\vskip -0.5mm \caption{Energy band dispersion in a shallow square
lattice potential with $t_1=-1,t_2=-0.6$ and $t_3=0$ (a), the real space order
parameter (b), and the checkerboard particle density (c) on a section of a $50\times50$ lattice. Here the
weak on-site repulsion is  $u_{00}=0.02$ , and the nearest-neighbor
 interaction $u_{01}=0.01$.}\label{fig:3}
\end{center}
\end{figure}

Some of our results are illustrated in Figs.
(\ref{fig:1},\ref{fig:2},\ref{fig:3}). Fig.(\ref{fig:1}a) represents
the single particle dispersion, $E({\bf k})$,  for a deep square
lattice potential, where $t_1$ is positive and $t_2$ and $t_3$ are
negligible. The level position in a single lattice valley is taken
as zero, $t(0)=0$, and all energies are measured in units of
$t_1=1$. The order parameter and the density are uniform in this
case, $f_{\bf m}=1$, and there is no phase transition with
increasing but moderate on-site repulsion, $u_{00} \lesssim 1$.
Effects driven by the strength of the repulsion will be typically
seen when $u_{00}\approx4.0$ or larger; these effects destroy the
homogeneity of the condensate.   Fig. (\ref{fig:1}b) shows an
example of such an inhomogeneous condensate. Fig (\ref{fig:1}c)
shows how the value of the on-site repulsion $u_{00}$ at which the
homogeneity is destroyed decreases with the value of $t_2$.  This
correlates with the increasing shallowness of the kinetic energy
dispersion as $t_2$ approaches $-0.5$, where the dispersion becomes
a pair of flat valley intercepting one another in a cruciform
centered at the $\Gamma$ point.   This homogeneous-inhomogeneous
transition occurs well below the critical value of $u_{00} \approx
16-23$ required for the superfluid-Mott insulator transition  on a
commensurate lattice
\cite{Krauth:1991,Krauth:1992,Cappello:2007,Capogrosso-Sansone:2008,Cappello:2008},
and so is a property of the superfluid state. There are different
irregular patterns in the inhomogeneous phase depending also on the
intersite interaction $u_{01}$, which are reminicent  but not
identical to  transformations of a homogeneous  Bose condensate into
a density wave superfluid \cite{aleran1981,kubotak} with a tendency
to the phase separation \cite{batscal}.

Can one increase the magnitude of $t_2$ further? In a conventional
optical lattice, one cannot, as a result of the textbook theorem
 to the effect that the ground state  wave-function of a single
particle must not contain any nodes \cite{Landau:1977,Courant:1989}.
This places limits on the values of hopping parameters that are
physically meaningful, since they cannot take values that give rise
to additional minima in the kinetic dispersion of the lowest band of
the Hamiltonian. However, there are a number of feasible situations
where the theorem is not applied due to  many-body effects
\cite{Kuklov:2006,Liu:2006}  or through the effects of internal
degrees of freedom \cite{alexandrov98,Larson:2008a}.
 Recent experimental \cite{Isacsson:2005,Muller:2007} and
theoretical work \cite{Stojanovic:2008} has also suggested that
condensation may be induced in an excited but  metastable p-band
state.    Also, this concern does not  arise with composite bosons
such as bipolarons \cite{alemot}, since the single polaron band,
where the condensation is occurring, is rarely the ground state band
due to the Pauli exclusion principle. In this case the hopping
parameters are determined by the symmetries of the wavefunctions of
two components of the boson, so that there is no constraint on
variation of any  hopping parameter.

With these considerations  we turn to the results presented in
Figures \ref{fig:2} and \ref{fig:3}.
Fig.(\ref{fig:2}) represents a shallow square lattice, where the
magnitude of $t_3$ is comparable with $t_1$. At some value of $t_3$
the minimum of the single-particle band shifts from the $\Gamma$
point with ${\bf k=0}$ to 4 finite wave-vectors within the first
Brillouin zone, ${\bf k}_{1}=k(1, 1)$, ${\bf k}_{2}=k(-1, 1)$, ${\bf
k}_{3}=k (-1, -1)$, and  ${\bf k}_{4}=k(1, -1)$ where Bose
condensation takes place, Fig.(\ref{fig:2}) (the value of $k$ is
calculated below). The four minima are degenerate.  A small
repulsion removes the degeneracy, so that the true condensate wave
function in the Wannier space is one of the superpositions
respecting the parity and time-reversal symmetry, $f_{\bf m}\propto
\cos ({\bf k_1 \cdot m})$, $f_{\bf m}\propto \cos ({\bf k_2 \cdot
m})$, or $f_{\bf m}\propto \cos ({\bf k_1 \cdot m})\pm \cos ({\bf
k_2 \cdot m})$, depending on the repulsion.  In our example with the
on-site repulsion only the order parameter and the density of
condensed bosons is striped along a diagonal direction,
Fig.(\ref{fig:2}). Here the nnn tunneling $t_3$ plays the role of a
parameter driving QPT. One can find its critical value, $t_{3c}$
corresponding to the transition from the uniform to the striped
condensate  by calculating the effective mass $m^*$  of a single
boson on the lattice. With the positive $t_1=1$, zero $t_2$ and a
negative $t_3$ the dispersion law is given by $E({\bf k}) = -2
[\cos(k_x ) +\cos(k_y )]-2t_3 [\cos(2k_x) +\cos(2k_y)]$, where the
lattice constant is taken as $a=1$. Expanding in powers of $k_{x,y}$
one obtains $E({\bf k}) \approx -4-4t_3 +\hbar^2k^2/2m^*$ with the
effective mass  $m^*= \hbar^2/(1+4t_3)$, so that QPT appears at
$t_{3c}=-0.25 $ when $m^*=\infty$. The QPT order parameter  is the
wave-vector (or inverse period of the modulation), which above
$t_{3c}$ is given by $k=\cos^{-1}(1/4|t_3|)$. It changes
continuously from zero at the transition up to $k=\pi/2$ at large
$|t_3|$, so that QPT is of the second order.  The transition is
driven entirely by the kinetic energy dispersion, rather than by the
repulsion,  so that it could be named a \emph{kinetic} quantum
phase transition (KQPT). A similar calculation for the case where
$t_2$ is varied and $t_3$ held fixed at zero shows that there is a
phase transition at $t_{2c}=-0.5$ to a state with 4 degenerate
minima located at ${\bf k}=(\pi,0),(-\pi,0),(0,\pi)$ and $(0,-\pi)$,
and is discussed in relation to Fig.(\ref{fig:3}).  Due to the
discontinuous nature of this phase transition, it is first order.

Quite generally, if the boson band dispersion has its minima at
finite ${\bf k}$  the Bose condensate is nonuniform.  Another
example is the center-of-mass band dispersion of a small bipolaron
composed of two holes on neighboring oxygen ions proposed as an
explanation of the unusual symmetry and checkerboard modulations of
the order parameter in cuprate superconductors \cite{alexandrov98}.
The nearest neighbor hopping integral is negative in this case due
to the $p$-symmetry of oxygen orbitals, so that the minima of the
bipolaronic band are found at the Brillouin zone boundaries.
Minimizing the energy functional (\ref{functional}) with negative
$t_1=-1$ yields different
 patterns of the order parameter and the density depending
on the interaction matrix elements and longer range hopping terms.
In this case we observe the relocation of 4 minima initially located
at ${\bf k}=(\pm\pi,\pm\pi)$ to new positions within the first
Brillouin zone, rather than a change from 1 to 4 minima. An
interesting example is the case of $t_2<-0.5$ and $t_3=0$, where a
weak nearest-neighbor repulsion $u_{10}$ can stabilize the d-wave
checkerboard order parameter (see Fig.(\ref{fig:3})) --  as
anticipated in Ref.\cite{alexandrov98} -- provided that $u_{00}$ is
not too strong.  Intriguingly, the same values of $t_2$, $t_3$ and
the repulsions {\em also} typically give a checkerboard state for
the positive $t_1=1$.

Cold atoms in optical lattices have provided an excellent tool for
investigating quantum phase transitions, but finding a reliable
diagnostic criterion for superfluidity is  not straightforward
\cite{diener,kato}. Our prediction of  QPTs in  dilute bosonic
superfluids, which is  strongly affected or even driven by the
kinetic energy dispersion,   opens up new perspectives on the unique
diagnostic criteria of superfluidity.  These states can most likley be detected
through the use of optical probe spectroscopy and/or phase sensitive experiments.

This work was supported  by EPSRC (UK) (grant no. EP/D035589/1).

\end{document}